\documentclass[conference,a4paper]{IEEEtran}

\usepackage{amsmath,amssymb,amsthm}

\def\Aut{\mathop{{\rm Aut}}}

\def\AGL{{\rm AGL}}
\def\GL{{\rm GL}}
\def\SL{{\rm SL}}
\def\PSL{{\rm PSL}}
\def\RM{{\rm RM}}
\def\CNOT{{\rm CNOT}}
\def\F{{\mathbb{F}}}
\def\Z{{\mathbb{Z}}}

\def\ket#1{|#1\rangle}
\def\wgt{{\rm wgt}}

\def\bm#1{\mathchoice{\mbox{\boldmath{$\displaystyle #1$}}}%
{\mbox{\boldmath{$\textstyle #1$}}}%
{\mbox{\boldmath{$\scriptstyle #1$}}}%
{\mbox{\boldmath{$\scriptscriptstyle #1$}}}}

\newtheorem{theorem}{Theorem}
\newtheorem{lemma}{Lemma}

\begin{document}

\title{Leveraging Automorphisms of Quantum Codes\\ for 
Fault-Tolerant Quantum Computation}

\author{
  \IEEEauthorblockN{Markus Grassl}
  \IEEEauthorblockA{Centre for Quantum Technologies\\
    National University of Singapore\\
    Email: Markus.Grassl@nus.edu.sg}
  \and
  \IEEEauthorblockN{Martin Roetteler}
  \IEEEauthorblockA{NEC Laboratories America\\
    Princeton, NJ, U.S.A.\\
    Email: mroetteler@nec-labs.com}
}

\maketitle

\begin{abstract}
Fault-tolerant quantum computation is a technique that is necessary to
build a scalable quantum computer from noisy physical building
blocks. Key for the implementation of fault-tolerant computations is
the ability to perform a universal set of quantum gates that act on
the code space of an underlying quantum code. To implement such a
universal gate set fault-tolerantly is an expensive task in terms of
physical operations, and any possible shortcut to save operations is
potentially beneficial and might lead to a reduction in overhead for
fault-tolerant computations. We show how the automorphism group of a
quantum code can be used to implement some operators on the encoded
quantum states in a fault-tolerant way by merely permuting the
physical qubits.  We derive conditions that a code has to satisfy in
order to have a large group of operations that can be implemented
transversally when combining transversal CNOT with automorphisms.  We
give several examples for quantum codes with large groups, including
codes with parameters $[\![8,3,3]\!]$, $[\![15,7,3]\!]$,
$[\![22,8,4]\!]$, and $[\![31,11,5]\!]$.
\end{abstract}

\section{Introduction}
Quantum error-correcting codes (QECC) are essential ingredients for
the realization of quantum computing devices.  In addition to the mere
error correction, it is also important that quantum operations can be
implemented in a fault-tolerant way, i.\,e., the operations preserve
the code space and if an operation fails, the errors remain local
\cite{Gottesman:97,NC:2000}. Several schemes are known for universal
fault-tolerant quantum computing, including schemes that are based on
distance-three codes \cite{ADP:2006} such as for instance the
concatenated Steane code \cite{Steane:2003,SDT:2007}, concatenated
error detecting codes \cite{Knill:2005}, or the Bacon-Shor codes
\cite{AC:2006}. Quite recently, the surface code---a stabilizer code
that exhibits one of the highest reported thresholds that exceed $1\%$
for a standard 2D lattice of physical qubits and independent
depolarizing noise---has gained a lot of attention
\cite{DKLP:2002,FMMC:2012}. So far, most of the schemes for
fault-tolerant quantum computing encode very few qubits per code
block; in the case of concatenated codes, typically QECCs are chosen
that encode only a single qubit per code block.

In this paper we present a general method that allows the
implementation of operations in a fault-tolerant manner for codes
encoding several qubits. Like in the single-qubit case, CSS codes
appear to be well suited for our methods, but they can be applied to
any stabilizer code. The basic idea is that code automorphisms can
give rise to non-trivial logical operations on the encoded quantum
information that can be executed by merely permuting, or what arguably
is simpler in a practical implementation, simply relabeling of the
physical qubits.  While such operations cannot {\em per se} give rise
to a universal gate set for which additional techniques such as state
distillation are essential, our construction can nevertheless lead to
operations that can be performed at basically zero cost.  This might
lead to overhead reductions, in particular for fault-tolerant quantum
computations on long block codes, provided they exhibit large
automorphism groups or automorphism groups with suitable structure.

\section{CSS Codes and their Automorphism Group}
First we consider the special case of CSS codes based on a classical
linear $C=[n,k_1,d_1]$ which is contained in its dual code
$C^\bot=[n,n-k_1,d_2]$. The (permutation) automorphism group $\Aut(C)$
of $C$ is the set of all permutations $\pi\in S_n$ that preserve the
code, i.\,e., (see also \cite{MS:77})
\begin{equation}
\forall\bm{c}\in C\colon\bm{c}^\pi=(c_{\pi(1)},\ldots,c_{\pi(n)})\in C.
\end{equation}
It turns out that $\Aut(C)=\Aut(C^\bot)$.
\begin{lemma}
Let $B=\{\bm{b}_1,\ldots,\bm{b}_{n-2k_1},\ldots,\bm{b}_{n-k_1}\}$ be a basis of
$C^\bot$ such that $B_0=\{\bm{b}_{n-2k_1},\ldots,\bm{b}_{n-k_1}\}$ is a basis of
$C$. With respect to the basis $B$, the automorphism group $\Aut(C)$
has a linear representation in the block-triangular form
\begin{alignat}{5}
\Aut(C)&\rightarrow \GL(n-k_1,2)\nonumber\\
\pi &\mapsto T(\pi)=
\left(
\begin{array}{c|c}
T_1(\pi) & T_2(\pi)\\
\hline
  0 & T_3(\pi)
\end{array}
\right).\label{eq:linear_action}
\end{alignat}
\end{lemma}
Recall that the basis states of the CSS code
$\mathcal{C}=[\![n,k,d]\!]$, where $k=n-2k_1$, based on the code $C$
are given by
\begin{equation}\label{CSS_state}
\ket{\psi_{\bm{v}}}=\frac{1}{\sqrt{|C|}}\sum_{\bm{c}\in C}\ket{\bm{c}+\bm{v}},
\end{equation}
where the vectors $\bm{v}=\sum_{i=1}^k\beta_i\bm{b}_i$ are
representatives of the cosets of $C$ in $C^\bot$. If we apply a
permutation $\pi\in\Aut(C)$ to the qudits of a basis state of the CSS
code, from eq. (\ref{eq:linear_action}) it follows that
\begin{equation}
\ket{\psi_{\bm{v}}}^\pi
=\frac{1}{\sqrt{|C|}}\sum_{\bm{c}\in C}\ket{\bm{c}^\pi+\bm{v}^\pi}
=\frac{1}{\sqrt{|C|}}\sum_{\bm{c}\in C}\ket{\bm{c}+\bm{v}'}
=\ket{\psi_{\bm{v}'}},
\end{equation}
where
\begin{equation}
\bm{v}'=\sum_{i=1}^k \beta_i'\bm{b}_i\quad\text{and}\quad
\beta_i'=\sum_{j=1}^k \bigl(T_1(\pi)\bigr)_{ij}\beta_j.
\end{equation}
Note that the basis state $\ket{\psi_{\bm{v}}}$ corresponds to the
encoding of the computational basis state $\ket{\bm{\beta}}$. Hence we
can label the basis states of the CSS code $\mathcal{C}$ by the vector
$\bm{\beta}=(\beta_1,\ldots,\beta_k)^T$.  Then we have
\begin{equation}
\ket{\bm{\beta}}^\pi=\ket{T_1(\pi)\bm{\beta}},
\end{equation}
i.\,e., the automorphism $\pi$ of the classical code $C$ gives rise to
a permutation of the basis states of the CSS code $\mathcal{C}$
corresponding to the linear transformation $T_1(\pi)$.  In summary we
have:
\begin{theorem}
Let $\mathcal{C}$ be a CSS code based on the classical code $C\le
C^\bot$. Then the automorphism $\pi\in\Aut(C)$ corresponds to the
linear operation $T_1(\pi)$ defined in eq. (\ref{eq:linear_action}) on
the logical basis states of $\mathcal{C}$.
\end{theorem}
In the general situation, a CSS code $\mathcal{C}$ is based on nested
classical codes $C_2\subset C_1$, and the basis states of
$\mathcal{C}$ correspond to the cosets of $C_2$ in $C_1$.  In general,
the automorphism groups $\Aut(C_1)$ and $\Aut(C_2)$ need not be equal.
However, when we consider their intersection, we obtain the following
result:
\begin{theorem}
Let $\mathcal{C}$ be a CSS code based on nested classical codes
$C_2\le C_1$. Then a joint automorphism $\pi\in\Aut(C_1)\cap\Aut(C_2)$
corresponds to a linear operation $T_1(\pi)$ on the logical basis
states of $\mathcal{C}$, defined analogously to
eq. (\ref{eq:linear_action}).
\end{theorem}

Note that these operations can be implemented by permuting the qubits
or just by relabeling them.  Below we will show that by a similar
argument, the (permutation) automorphism group of an additive code
corresponding to a stabilizer code gives rise to symplectic operations
on the logical operators of the stabilizer code. We would also like to
point out that while automorphism groups of additive codes have been
investigated before, see e.\,g., \cite{CRSS:98,Rains:99,ZCC:2011}, the
idea to leverage automorphisms to perform large sets of encoded
logical operations does not seem to have been investigated
much.\footnote{However, we would like to point out that the
  automorphism group of the quantum Hamming code of length $15$ was
  used to aid fault-tolerant quantum computation in a talk given by
  J.~Harrington at the QEC 2011 conference.}

\section{Combining Automorphisms and Transversal Operations}
For CSS codes, applying the controlled-NOT (CNOT) operation
transversally is an operation preserving the space of two copies of
the code. More precisely, we have
\begin{equation}
\CNOT^{\otimes n}\bigl(\ket{\psi_{\bm{v}_1}}\ket{\psi_{\bm{v}_2}}\bigr)
=\ket{\psi_{\bm{v}_1}}\ket{\psi_{\bm{v}_1+\bm{v}_2}},
\end{equation}
where $\CNOT^{\otimes n}$ should be understood as applying
$\CNOT$-gates to the corresponding qudits in both code blocks. In terms
of the encoded basis states, we have
\begin{equation}
\CNOT^{\otimes n}\bigl(\ket{\bm{\beta_1}}\ket{\bm{\beta_2}}\bigr)
=\ket{\bm{\beta_1}}\ket{\bm{\beta_1+\beta_2}},
\end{equation}
i.\,e., the transversal CNOT corresponds to the linear $2k\times 2k$
matrix
\begin{equation}
\left(\def\arraystretch{1.1}
\begin{array}{c|c}
I & 0\\
\hline
I & I
\end{array}
\right).
\end{equation}
In the following we assume that the CNOT-gates can not only be applied
to the corresponding pairs of qudits in each code block, but between
any pair of qudits.  Then we can combine the operations on the code
arising from the automorphism group of the underlying classical code
and the transversal CNOT.
\begin{theorem}
Given a CSS code $\mathcal{C}=[\![n,k,d]\!]$ derived from a linear
code $C\le C^\bot$ with automorphism group $\Aut(C)$, one can realize
the following group $G_{12}$ of linear transformations on $2k$ encoded
qudits in a fault-tolerant manner:

\begin{equation}
\hspace*{-4cm}
G_{12}=\left\langle
\left(\def\arraystretch{1.1}
\begin{array}{c|c}
I & 0\\
\hline
I & I
\end{array}
\right),
\left(\def\arraystretch{1.1}
\begin{array}{c|c}
I & I\\
\hline
0 & I
\end{array}
\right),\right.
\end{equation}
\begin{equation*}
\left.
\left(\def\arraystretch{1.1}
\begin{array}{c|c}
T_1(\pi_1) & 0\\
\hline
0 & T_1(\pi_2)
\end{array}
\right)\colon
\pi_1,\pi_2\in\Aut(C)
\right\rangle.
\end{equation*}
\end{theorem}
The first two generators of $G_{12}$ are the transversal CNOT with all
controls in the first or second code block, respectively.  While we
cannot make a general statement about the relation between the
automorphism group $\Aut(C)$ and the group $G_{12}$, we have the
following observation.
\begin{lemma}
The group $G_{12}$ contains all matrices of the form
\begin{equation}
\left(\def\arraystretch{1.1}
\begin{array}{c|c}
I & A\\
\hline
0 & I
\end{array}
\right)
\quad\text{and}\quad
\left(\def\arraystretch{1.1}
\begin{array}{c|c}
I & 0\\
\hline
A & I
\end{array}
\right),
\end{equation}
where $A$ is an arbitrary element of the $\Z$-algebra generated by the
matrices $T_1(\pi_j)$, i.\,e.,
\begin{equation}
A=\sum_{\pi\in\Aut(C)} \alpha_\pi T_1(\pi),\qquad \alpha_\pi\in\Z.
\end{equation}
Hence we can in particular realize transformations of the form
\begin{equation}
\ket{\bm{\beta}_1}\ket{\bm{\beta}_2}\mapsto\ket{\bm{\beta}_1}\ket{A\bm{\beta}_1+\bm{\beta_2}}.
\end{equation}
\end{lemma}
\begin{IEEEproof}
First, note that
\begin{equation}
\left(\def\arraystretch{1.1}
\begin{array}{c|c}
T_1(\pi) & 0\\
\hline
0 & I
\end{array}
\right)
\left(\def\arraystretch{1.1}
\begin{array}{c|c}
I & I\\
\hline
0 & I
\end{array}
\right)
\left(\def\arraystretch{1.1}
\begin{array}{c|c}
T_1(\pi) & 0\\
\hline
0 & I
\end{array}
\right)^{-1}
\end{equation}
\begin{equation*}
=
\left(\def\arraystretch{1.1}
\begin{array}{c|c}
I & T_1(\pi)\\
\hline
0 & I
\end{array}
\right).
\end{equation*}
The products of these matrices and their inverses yield arbitrary
integer linear combinations of the matrices $T_1(\pi_j)$ in the
upper right block.  The result for lower-triangular block matrices
follows analogously.
\end{IEEEproof}

\begin{theorem}\label{thm:gens_SL}
Assume that the group $G_{12}$ contains all matrices of the form
\begin{equation}
\left\{
\left(
\begin{array}{c|c}
I & A\\
\hline
0 & I
\end{array}
\right)
\colon A\in M_{n\times n}(\F_q)\right\}
\quad\text{and}\quad
\end{equation}
\begin{equation*}
\left\{\left(
\begin{array}{c|c}
I & 0\\
\hline
B & I
\end{array}
\right)
\colon B\in M_{n\times n}(\F_q)\right\},
\end{equation*}
where $A,B\in M_{n\times n}(\F_q)$ are arbitrary matrices of the algebra
of $n\times n$ matrices over the field $\F_q$.  Then $G_{12}= \SL_{2n}(\F_q)$.
\end{theorem}
\begin{IEEEproof}
Let $E_{i,j}$ denote the $n\times n$ matrix which has the entry $1$ in
row $i$ and column $j$, and is zero elsewhere.  By assumption, the
group $G_{12}$ contains the following two matrices:
\begin{equation}\label{eq:transvections0}
M_1=\left(
\begin{array}{c|c}
I & \alpha E_{i,j}\\
\hline
0 & I
\end{array}
\right)
\quad\text{and}\quad
M_2=\left(
\begin{array}{c|c}
I & 0\\
\hline
\beta E_{j,k} & I
\end{array}
\right)
\end{equation}
with $i\ne k$.  
We compute
\begin{alignat}{5}
&M_2^{-1}M_1 M_2 M_1^{-1}=\left(
\begin{array}{c|c}
I+\alpha\beta E_{i,k}&0\\
\hline
0 & I
\end{array}
\right).\label{eq:transvections}
\end{alignat}

By symmetry, we also get the same type of matrices in the lower right
block, and in summary all elementary transvections with identity on
the diagonal and a single non-zero off-diagonal entry.  Furthermore,
for $t\ne 0$ we obtain the following factorizations of diagonal
matrices:
\begin{alignat}{5}
&\kern-2ex\left(\arraycolsep0.75\arraycolsep
\begin{array}{cc|cc}
 t &   0 & 0 & 0\\
 0 & 1/t & 0 & 0\\\hline
 0 &   0 & 1 & 0\\
 0 &   0 & 0 & 1
\end{array}
\right)=\nonumber\\
&\left(\arraycolsep0.75\arraycolsep
\begin{array}{cc|cc}
 1 &      0 & 0 & 0\\
 0 &      1 & 0 & 0\\\hline
 0 &      0 & 1 & 0\\
 0 &(t-1)/t & 0 & 1
\end{array}
\right)
\left(\arraycolsep0.75\arraycolsep
\begin{array}{cc|cc}
 1 & 0 & 0 & 1\\
 0 & 1 & 0 &-1\\\hline
 0 & 0 & 1 & 0\\
 0 & 0 & 0 & 1
\end{array}
\right)\nonumber\\
&\times
\left(\arraycolsep0.75\arraycolsep
\begin{array}{cc|cc}
      1 & 0 & 0 & 0\\
      0 & 1 & 0 & 0\\\hline
      0 & 0 & 1 & 0\\
(1-t)/t & 0 & 0 & 1\\
\end{array}
\right)
\left(\arraycolsep0.75\arraycolsep
\begin{array}{cc|cc}
 1 & 0 & 0 &-t\\
 0 & 1 & 0 & 0\\\hline
 0 & 0 & 1 & 0\\
 0 & 0 & 0 & 1\\
\end{array}
\right)\nonumber\\
&\times\left(\arraycolsep0.75\arraycolsep
\begin{array}{cc|cc}
        1 &      0 & 0 & 0\\
        0 &      1 & 0 & 0\\\hline
        0 &      0 & 1 & 0\\
(t-1)/t^2 &(1-t)/t & 0 & 1
\end{array}
\right)
\left(\arraycolsep0.75\arraycolsep
\begin{array}{cc|cc}
1 &0 &0 &0\\
0 &1 &0 &1\\\hline
0 &0 &1 &0\\
0 &0 &0 &1
\end{array}
\right)\label{eq:diagonal_one_block}
\end{alignat}
and
\begin{alignat}{5}
\kern-2ex\left(\arraycolsep0.75\arraycolsep
\begin{array}{cc|cc}
 t & 0 &   0 & 0\\
 0 & 1 &   0 & 0\\\hline
 0 & 0 & 1/t & 0\\
 0 & 0 &   0 & 1
\end{array}
\right)=
&\left(\arraycolsep0.75\arraycolsep
\begin{array}{cc|cc}
1 & 0 & t-1 & 0\\
0 & 1 &   0 & 0\\\hline
0 & 0 &   1 & 0\\
0 & 0 &   0 & 1
\end{array}
\right)
\left(\arraycolsep0.75\arraycolsep
\begin{array}{cc|cc}
 1 & 0 & 0 & 0\\
 0 & 1 & 0 & 0\\\hline
 1 & 0 & 1 & 0\\
 0 & 0 & 0 & 1
\end{array}
\right)\nonumber\\
&\hspace*{-2cm}\times\left(\arraycolsep0.75\arraycolsep
\begin{array}{cc|cc}
 1 & 0 & (1-t)/t & 0\\
 0 & 1 &       0 & 0\\\hline
 0 & 0 &       1 & 0\\
 0 & 0 &       0 & 1
\end{array}
\right)
\left(\arraycolsep0.75\arraycolsep
\begin{array}{cc|cc}
  1 & 0 & 0 & 0\\
  0 & 1 & 0 & 0\\\hline
 -t & 0 & 1 & 0\\
  0 & 0 & 0 & 1
\end{array}
\right).\label{eq:diagonal_two_blocks}
\end{alignat}
The matrices of the form (\ref{eq:diagonal_one_block}) and
(\ref{eq:diagonal_two_blocks}) generate all diagonal matrices with
unit determinant.  Together with the transvections in
(\ref{eq:transvections0}) and (\ref{eq:transvections}), they generate
the full special linear group $\SL_2(\F_q)$.
\end{IEEEproof}

\section{Examples}
Good candidates for this construction are codes with large
automorphism group or automorphism groups for which the representation
given by $T_1(\pi)$ is irreducible or has only a few irreducible
components of large dimension.  Among those, Reed-Muller codes and
cyclic codes are promising candidates.
\subsection{CSS code $[\![15,7,3]\!]$}
The $4$th-order binary Hamming code has parameters $[15,11,3]$ and
contains its dual code $C=[15,4,8]$. The automorphism group of $C$ is
isomorphic to the alternating group $A_8$ of order $21600$.

The linear action on the $7$ logical qubits is given by the group
\begin{equation}
G_1=\left\langle\arraycolsep0.5\arraycolsep
\begin{pmatrix}
1&0&0&1&1&0&1\\
1&1&0&0&1&0&0\\
1&1&1&0&1&1&1\\
1&1&0&0&0&1&0\\
0&1&0&0&1&0&1\\
0&0&0&1&1&0&1\\
1&1&0&0&1&1&0
\end{pmatrix}
,
\begin{pmatrix}
1&0&1&0&0&1&0\\
1&1&1&1&1&0&0\\
0&1&1&0&1&1&0\\
0&1&0&1&0&1&1\\
1&0&0&1&0&0&0\\
0&1&0&0&1&1&0\\
0&1&1&1&1&0&1
\end{pmatrix}
\right\rangle.
\end{equation}
Combining the group $G_1\times G_1$ with the transversal CNOTs, we get
the group $G_{12}\cong \SL(12,2)\times\SL(2,2)$ with more than
$2^{144}$ elements.  What is even more, the block-diagonal subgroup
$\tilde{G}_1$ of $G_{12}$ that acts trivially on the second code block
is isomorphic to the group $\SL(6,2)$.

A closer inspection shows that the Hamming code contains the all-one
vector which corresponds to the logical operator $X^{\otimes 15}$ on
the code.  Both are invariant under permutations.  Hence on the
subcode $[\![15,6,3]\!]$ of the original code, which is obtained by
removing the all-one vector from the Hamming code, we can realize the
full linear group $\SL(6,2)$ on the encoded states as well as the full
linear group $\SL(12,2)$ on pairs of encoded states.

\subsection{CSS code $[\![31,11,5]\!]$}
The BCH code with parameters $[31,21,5]$ contains its dual
$C=[31,10,12]$. The resulting CSS code has parameters
$\mathcal{C}=[\![31,11,5]\!]$. The automorphism group of $C$ is a
group $G_1$ of order $155$ isomorphic to $C_{31}\rtimes C_5$. However,
when combining $G_1\times G_1$ with the transversal CNOTs, we obtain
the group $G_{12}$ isomorphic to
$\SL(10,2)\times\SL(10,2)\times\SL(2,2)$ with more than $2^{199}$
elements. Restricted to one code block, we get the group
$\tilde{G}_1\cong\SL(5,2)\times\SL(5,2)$. Similar as for the CSS code
$[\![15,7,3]\!]$, we find that the spaces of dimension $5$, $5$, and
$1$ stabilized by the code correspond to cyclic subcodes lying between
the code $C$ and $C^\bot$.  On each of the subspaces, we can realize
the full linear group, despite the fact that the automorphism group
$\Aut(C)$ is relatively small.

\subsection{CSS code $[\![22,8,4]\!]$}
The classical self-orthogonal code $C=[22,7,8]$
generated by
\begin{equation}
G=\left(\arraycolsep0.4\arraycolsep
\begin{array}{*{22}{c}}
1&0&0&0&1&0&0&0&0&1&1&0&0&0&1&0&1&0&1&0&1&0\\
0&1&0&0&1&0&0&0&0&1&1&0&1&0&0&1&0&1&0&1&1&1\\
0&0&1&0&1&0&0&0&0&1&0&1&0&0&1&1&0&1&1&0&0&0\\
0&0&0&1&0&0&0&1&0&0&1&0&0&0&0&1&1&0&1&0&1&1\\
0&0&0&0&0&1&0&1&0&1&0&1&1&1&1&1&1&0&0&0&1&0\\
0&0&0&0&0&0&1&1&0&0&1&0&0&1&1&0&1&0&1&1&0&0\\
0&0&0&0&0&0&0&0&1&1&1&1&1&1&1&1&1&1&1&1&1&1
\end{array}
\right)
\end{equation}
is contained in its dual $C^\bot=[22,15,4]$.  Hence we obtain a CSS
code $\mathcal{C}=[\![22,8,4]\!]$.  The automorphism group of $C$ has
order 336 and is isomorphic to a semi-direct product of $\PSL(2,7)$ and
$Z_2$.  Although the group is relatively small, the action on the
space of $8$ logical qubits is an irreducible matrix group
\begin{equation}
G_1=\left\langle\arraycolsep0.5\arraycolsep
\begin{pmatrix}
1&1&0&1&0&1&1&0\\
0&1&1&1&1&1&0&0\\
0&1&1&0&1&1&0&1\\
1&1&1&0&0&0&0&0\\
1&0&1&0&1&1&0&0\\
1&1&0&1&1&1&0&1\\
0&0&1&0&0&1&0&0\\
1&0&1&0&0&1&1&0
\end{pmatrix}
,
\begin{pmatrix}
1&1&1&0&0&0&0&1\\
0&1&0&1&1&0&1&0\\
0&1&0&0&1&0&1&1\\
1&0&1&0&1&0&0&1\\
1&1&0&0&0&0&1&1\\
1&1&1&0&1&1&0&1\\
0&0&0&0&0&0&1&0\\
1&0&1&1&0&0&0&0
\end{pmatrix}
\right\rangle.
\end{equation}
The matrices in $G_1$ span the full space of binary $8\times 8$
matrices.  Hence by Theorem \ref{thm:gens_SL}, combining the group
$G_1\times G_1$ with the transversal CNOTs we get the maximal possible
group $\SL(16,2)$.

\subsection{Stabilizer code $[\![8,3,3]\!]$}
There is a stabilizer code $\mathcal{C}=[\![8,3,3]\!]$ whose five
generators of the stabilizer, the three logical $X$-operators, and the
three logical $Z$-operators correspond to the following vectors (top
to down, respectively) over $GF(4)$:
\begin{equation}
\left(\let\w\omega
\begin{array}{cccccccc}
 1 &  0 & \w &   0 &\w^2 &  \w &   1 &\w^2\\
\w &  0 & \w &   1 &   0 &\w^2 &\w^2 &  1\\
 0 &  1 & \w &  \w &\w^2 &   1 &\w^2 &  0\\
 0 & \w &  0 &\w^2 &  \w &   1 &   1 &\w^2\\
 0 &  0 &  1 &\w^2 &   1 &  \w &\w^2 &  \w\\
\hline
0 &  0 & \w &  1 & \w &\w^2 &  1 &\w^2\\
0 &  0 &  0 & \w &  0 &  \w & \w &  \w\\
0 &  0 & \w &  0 & \w &  \w &  0 &  \w\\
\hline
0 &  0 &  0 &  1 & \w &  0 &\w^2 &  0\\
0 &  0 &  0 &  0 &  1 &  0 &  \w &\w^2\\
0 &  0 & \w &  0 &  0 &  0 &\w^2 &  1
\end{array}
\right)
\end{equation}
Here $\omega\in GF(4)$ obeys the relation $\omega^2=\omega+1$, and
Pauli matrices $X$, $Y$, and $Z$ correspond to $1$, $\omega^2$, and
$\omega$, respectively. The permutation automorphism group of
$\mathcal{C}$ is isomorphic to the group $\rm{AGL}(1,8)$ of order
$56$. On the symplectic space of the logical operators of
$\mathcal{C}$, we have the following matrix representation of
$\Aut(\mathcal{C})$:
\[\arraycolsep0.75\arraycolsep
G_1=\left\langle
\left(
\begin{array}{ccc|ccc}
0&1&0&0&0&0\\
0&0&1&0&0&0\\
1&0&1&0&0&0\\
\hline
1&0&0&0&1&0\\
0&1&1&1&0&1\\
0&1&0&1&0&0
\end{array}
\right),
\left(
\begin{array}{ccc|ccc}
1&0&0&0&0&0\\
0&1&0&0&0&0\\
0&0&1&0&0&0\\
\hline
0&1&0&1&0&0\\
1&0&0&0&1&0\\
0&0&0&0&0&1
\end{array}
\right)
\right\rangle
\]
Note that with this choice of logical operators, the space
corresponding to the logical $X$-operators is preserved.  

Unlike the situation for CSS codes, the transversal CNOT-gate does not
preserve stabilizer codes in general. So we have to look for
stabilizer codes which have a larger symmetry group. Additionally, we
may consider the automorphism group including local Clifford
operations as well.

\section{Code Families}
We briefly discuss the situation for CSS codes based on Reed-Muller
codes or cyclic codes. 

\subsection{Reed-Muller codes}
Recall that the $r$-th order binary Reed-Muller code $\RM(r,m)$ of
length $n=2^m$, for $0\le r\le m$ is obtained by the evaluation of all
Boolean functions in $m$ variables of maximal degree $r$ (see, e.\,g.,
\cite{MS:77}).  The automorphism group of $\RM(r,m)$ contains the
group $\AGL(m,2)$ of all affine transformations on $\F_2^m$.  As
affine transformations preserve the degree of Boolean functions, it
follows that the automorphism group also preserves the cosets of
$\RM(r,m)$ in $\RM(r+1,m)$.  Hence, if a CSS code is based on the
nested codes $\RM(r,m)\subset RM(r+s,m)$, the action of $\AGL(m,2)$ on
the CSS code will not mix the blocks of logical qubits corresponding
to homogeneous Boolean functions of fixed degree.  Additional
automorphisms or other techniques are needed to implement operations
between the blocks.

\subsection{Cyclic codes}
Recall that every linear binary cyclic code of odd length $n$ can be
uniquely described by a generator polynomial $g(X)$ that divides
$X^n-1$.  Given two nested cyclic codes $C_2=[n,k_2]\subset
C_1=[n,k_1]$, their generator polynomials obey the relation
$g_2(X)=g_1(X) h(X)$, where $h(X)$ is some factor of $X^n-1$ of degree
$k_1-k_2$.  Assume that the polynomial $h(X)$ has irreducible factors
$h_i(X)$ of degree $\delta_i$, respectively.  Then the coset space
$C_1/C_2$ can be decomposed into spaces of dimension $\delta_i$ which
are preserved by the action of the cyclic group $Z_n$ of order $n$.
In turn, for a cyclic CSS code based on $C_2\subset C_1$, the cyclic
shift gives rise to operations on blocks with $\delta_i$ logical
qubits.  If $n<\delta_i^2$, the matrices corresponding to the action
on these blocks do not generate the full algebra of
$\delta_i\times\delta_i$ matrices.  Hence we cannot apply Theorem
\ref{thm:gens_SL}, and it is not clear whether we can implement the
full group of linear transformations on that block with $\delta_i$
logical qubits.  Of course the situation changes when there are more
automorphisms than just the cyclic shift.

\section{Towards the full Clifford Group}\label{sec:Clifford}
When the conditions in Theorem \ref{thm:gens_SL} are met, we can
implement all linear transformations on a single block of $k$ logical
qudits as well as on any number of such blocks.  Using tensor products
of local $X$-operations corresponding to coset representatives
$\bm{v}$, we can implement affine shifts on the logical qudits, and
hence all affine transformations.

If a CSS code is based on a classical self-orthogonal code $C\le
C^\bot$, we can apply a local Fourier transformation transversally on
all qudits, resulting in a simultaneous Fourier transformation on all
logical qudits.  This operation will interchange the role of the
logical $X$- and $Z$-operations.  In order to implement all Clifford
operations on the logical qudits, additional transformations that mix
$X$- and $Z$-operations are required.

If the CSS code is based on a doubly-even binary code, applying the
local transformation $P={\rm diag}(1,i)$, where $i^2=-1$,
transversally induces an operation on the code states
(\ref{CSS_state}) given by
\begin{equation}
P^{\otimes n}\ket{\psi_{\bm{v}}}=i^{\wgt(\bm v)}\ket{\psi_{\bm{v}}}.
\end{equation}
Hence depending on $\wgt(\bm{v})\bmod 4$, different powers of $P$ are
applied to the corresponding logical state.  The favorable situation
is when we indeed have a different action on the logical qubits.  In
that case, the combination with permutations of the logical qubits
(which are in particular linear transformations) yields a larger group
of transformations on the logical qudits.  The very group, however,
depends on the particular code.

\enlargethispage{-4.5cm} 

\section{Conclusions}

We proposed a general method that allows the implementation of
operations in a fault-tolerant manner for codes encoding several
qubits. In Theorem \ref{thm:gens_SL} we presented a sufficient
condition on the automorphism group of a quantum code such that all
linear transformations on the logical qubits can be implemented by
permutations of the qubits and transversal CNOT operations. We applied
this to a set of examples, including quantum codes with parameters
$[\![8,3,3]\!]$, $[\![15,7,3]\!]$, $[\![22,8,4]\!]$, and
$[\![31,11,5]\!]$.  Furthermore, we discussed the prospects for
applying this framework to infinite families of quantum block codes,
such as the Reed-Muller codes and cyclic codes. There are several open
questions that are implied by these observations: (i) Can we find more
examples of quantum codes for which the complete set of linear
transformations can be implemented following Theorem
\ref{thm:gens_SL}? In particular, it would be interesting to know if
code families with this property exist that are asymptotically good.
(ii) Can we find codes---or families of codes---for which we can
implement not only all linear transformations, but the full Clifford
group on $k$ logical qudits extending the results shown here?

\section*{Acknowledgments}
Supported in part by the Intelligence Advanced Research Projects
Activity (IARPA) via Department of Interior National Business Center
Contract number DllPC20l66. The U.S.  Government is authorized to
reproduce and distribute reprints for Governmental purposes
notwithstanding any copyright annotation thereon. Disclaimer: The
views and conclusions contained herein are those of the authors and
should not be interpreted as necessarily representing the official
policies or endorsements, either expressed or implied, of IARPA,
DoI/NBC or the U.S. Government.

The Centre for Quantum Technologies (CQT) is a Research Centre of
Excellence funded by the Ministry of Education and the National
Research Foundation of Singapore.


\begin{thebibliography}{10}
\providecommand{\url}[1]{#1}
\csname url@rmstyle\endcsname
\providecommand{\newblock}{\relax}
\providecommand{\bibinfo}[2]{#2}
\providecommand\BIBentrySTDinterwordspacing{\spaceskip=0pt\relax}
\providecommand\BIBentryALTinterwordstretchfactor{4}
\providecommand\BIBentryALTinterwordspacing{\spaceskip=\fontdimen2\font plus
\BIBentryALTinterwordstretchfactor\fontdimen3\font minus
  \fontdimen4\font\relax}
\providecommand\BIBforeignlanguage[2]{{%
\expandafter\ifx\csname l@#1\endcsname\relax
\typeout{** WARNING: IEEEtran.bst: No hyphenation pattern has been}%
\typeout{** loaded for the language `#1'. Using the pattern for}%
\typeout{** the default language instead.}%
\else
\language=\csname l@#1\endcsname
\fi
#2}}

\bibitem{Gottesman:97}
D.~Gottesman, ``Stabilizer codes and quantum error correction,'' Ph.D.
  dissertation, Caltech, 1997, see also: arXiv preprint quant-ph/9705052.

\bibitem{NC:2000}
M.~Nielsen and I.~Chuang, \emph{Quantum Computation and Quantum
  Information}.\hskip 1em plus 0.5em minus 0.4em\relax Cambridge, UK: Cambridge
  University Press, 2000.

\bibitem{ADP:2006}
P.~Aliferis, D.~Gottesman, and J.~Preskill, ``Quantum accuracy threshold for
  concatenated distance-3 codes,'' \emph{Quant. Information and Computation},
  vol.~6, no.~2, pp. 97--165, 2006.

\bibitem{Steane:2003}
A.~Steane, ``Overhead and noise threshold of fault-tolerant quantum error
  correction,'' \emph{Phys.~Rev.~A}, vol.~68, p. 042322, 2003.

\bibitem{SDT:2007}
K.~M. Svore, D.~P. DiVincenzo, and B.~M. Terhal, ``Noise threshold for a
  fault-tolerant two-dimensional lattice architecture,'' \emph{Quant.
  Information and Computation}, vol.~7, no.~4, pp. 297--318, 2007.

\bibitem{Knill:2005}
E.~Knill, ``Quantum computing with realistically noisy devices,''
  \emph{Nature}, vol. 434, pp. 39--44, 2005.

\bibitem{AC:2006}
P.~Aliferis and A.~Cross, ``{Subsystem fault tolerance with the Bacon-Shor
  code},'' \emph{Phys. Rev. Lett.}, vol.~98, p. 220502, 2006.

\bibitem{DKLP:2002}
E.~Dennis, A.~Alexei~Kitaev, A.~Landahl, and J.~Preskill, ``Topological quantum
  memory,'' \emph{J. Math. Phys.}, vol.~43, pp. 4452--4505, 2002.

\bibitem{FMMC:2012}
A.~G. Fowler, M.~Mariantoni, J.~M. Martinis, and A.~N. Cleland, ``Surface
  codes: Towards practical large-scale quantum computation,''
  \emph{Phys.~Rev.~A}, vol.~86, p. 032324, 2012.

\bibitem{MS:77}
F.~J. MacWilliams and N.~J.~A. Sloane, \emph{The Theory of Error--Correcting
  Codes}.\hskip 1em plus 0.5em minus 0.4em\relax Amsterdam: North--Holland,
  1977.

\bibitem{CRSS:98}
A.~R. Calderbank, E.~M. Rains, P.~W. Shor, and N.~J.~A. Sloane, ``Quantum error
  correction via codes over {GF}(4),'' \emph{IEEE Trans. Inform. Theory},
  vol.~44, pp. 1369--1387, 1998.

\bibitem{Rains:99}
E.~M. Rains, ``Quantum codes of minimum distance two,'' \emph{IEEE Trans.
  Inform. Theory}, vol.~45, no.~1, pp. 266--271, 1999.

\bibitem{ZCC:2011}
B.~Zeng, A.~Cross, and I.~L. Chuang, ``Transversality versus universality for
  additive quantum codes,'' \emph{IEEE Trans. Inform. Theory}, vol.~57, no.~9,
  pp. 6272--6284, 2011.

\end{thebibliography}
\end{document}